\documentstyle[12pt]{article}
\newcommand{\beq}{\begin{equation}}
\newcommand{\eeq}{\end{equation}}
\newcommand{\beqa}{\begin{eqnarray}}
\newcommand{\eeqa}{\end{eqnarray}}
\newcommand{\beqan}{\begin{eqnarray*}}
\newcommand{\eeqan}{\end{eqnarray*}}
\newcommand{\ba}{\begin{array}}
\newcommand{\ea}{\end{array}}
\newcommand{\no}{\nonumber}

\newcommand{\ol}{\overline}
\newcommand{\ra}{\rightarrow}

\newcommand{\ve}{\varepsilon}
\newcommand{\vp}{\varphi}

\newcommand{\dg}{\dagger}
\newcommand{\wt}{\widetilde}
\newcommand{\wh}{\widehat}

\newcommand{\A}{{\cal A}}

\newcommand{\D}{{\cal D}}

\newcommand{\cL}{{\cal L}}
\newcommand{\M}{{\cal M}}

\newcommand{\cO}{{\cal O}}

\newcommand{\V}{{\cal V}}

\newcommand{\dfrac}{\displaystyle \frac}

\newcommand{\co}{\; \; ,}
\newcommand{\nn}{\nonumber \\}
\newcommand{\scs}{\co \;}

\newcommand{\bea}{\begin{eqnarray}}
\newcommand{\eea}{\end{eqnarray}}
\newcommand{\del}{\partial}

\normalsize
\begin{document}
\begin{titlepage}
\begin{flushright}
CERN-TH/98-231\\
UWThPh-1998-42\\
\end{flushright}
\vspace{2.5cm}
\begin{center}
{\Large \bf  The Super-Heat-Kernel Expansion and the Renormalization
of the Pion--Nucleon Interaction*}\\[40pt]
H. Neufeld

\vspace{1cm}
 CERN, CH-1211 Geneva 23, Switzerland \\[10pt]

                        and \\[10pt]  

Institut f\"ur Theoretische Physik, Universit\"at
Wien\\ Boltzmanngasse 5, A-1090 Vienna, Austria

\vfill
{\bf Abstract} \\
\end{center}
\noindent
A recently proposed super-heat-kernel technique is applied to $SU(2)_L
\times SU(2)_R$ heavy baryon chiral perturbation theory. A previous 
result for the one-loop divergences of the pion--nucleon
system to $\cO(p^3)$ is confirmed, giving at the same time an impressive
demonstration of the efficiency of the new method. The cumbersome and tedious
calculations of the conventional approach are now reduced to a few
simple algebraic manipulations. The present computational scheme is not 
restricted to chiral perturbation theory, but can easily be applied or
extended to any (in general non-renormalizable) theory with boson--fermion
interactions.

\vfill
 
\begin{flushleft} CERN-TH/98-231\\
July 1998\\ 
\end{flushleft}

\noindent * Work supported in part by  TMR, EC-Contract No. ERBFMRX-CT980169 
(EURODA$\Phi$NE) 

\end{titlepage}
\section{Introduction}
\label{sec: Introduction}
\renewcommand{\theequation}{\arabic{section}.\arabic{equation}}
\setcounter{equation}{0}

The modern treatment of the pion--nucleon system as an effective field
theory of the standard model was pioneered by Gasser, Sainio, \v{S}varc 
\cite{GSS88} and Krause \cite{Krause} who formulated the ``relativistic''
version of baryon chiral
perturbation theory. It was then shown by Jenkins and Manohar \cite{JM91}
that the methods of heavy quark effective theory \cite{HQET} allow for a
systematic low-energy expansion of baryonic Green functions in complete
analogy to the meson sector. The latter approach is usually called heavy
baryon chiral perturbation theory.

Applications of this effective field theory beyond the tree level require
the knowledge of the divergences generated by one-loop graphs. For the
pion--nucleon interaction in the heavy mass expansion, the full list of
one-loop divergences to $\cO(p^3)$ has been worked out by Ecker
\cite{Eck94}. This
analysis was then extended to the three-flavour case by M\"uller and
Mei\ss ner \cite{MM96}. In these papers, the bosonic loop 
and the mixed loop (boson and
fermion lines in the loop) were treated separately. This required a
cumbersome investigation of the
singular behaviour of products of
propagators, because the mixed loop does not have the form of a
determinant, like the purely bosonic or fermionic loops.

To overcome these difficulties, we have recently developed \cite{NGE98} a
method where bosons and fermions are treated on an equal footing.
Employing the notions of supermatrices, superdeterminants and supertraces
\cite{ANZ,berezin}, we have constructed a super-heat-kernel
representation for the one-loop functional of a boson-fermion system. In
this approach, the determination of the one-loop divergences is reduced to
simple matrix manipulations, in complete analogy to the familiar
heat-kernel expansion technique for bosonic or fermionic loops.    

The present paper is organized as follows: in Sect. \ref{sec: SHK} I
briefly
review the super-heat-kernel method.
In contrast to the Euclidean space formulation used in \cite{NGE98}, the
presentation in this work refers to Minkowski space throughout. In Sect.
\ref{sec: SFI} 
the super-heat-kernel formalism is applied to a rather general class of
scalar--heavy fermion interactions (including, of course, heavy baryon
chiral perturbation theory). The one-loop divergences to second order in
the fermion fields are given explicitly. These results are then
specialized to the two-flavour version of heavy baryon chiral perturbation
theory in Sect. \ref{sec: renorm}. My conclusions, together with an
outlook to possible extensions of the present work, are summarized in
Sect. \ref{sec: Conclusions}. Several momentum-space integrals are
collected in the Appendix.

\section{Super-Heat-Kernel}
\label{sec: SHK}
\renewcommand{\theequation}{\arabic{section}.\arabic{equation}}
\setcounter{equation}{0}

Let us consider a general action
\beq
\label{action}
S[\vp,\psi,\ol \psi] = \int d^dx ~ \cL(\vp,\psi,\ol \psi)
\eeq
for $n_B$ real scalar fields $\vp_i$
and $n_F$ spin 1/2 fields $\psi_a$. Anticipating the later use of
dimensional regularization, I am starting in $d$-dimensional Minkowski
space. To construct the generating
functional $Z$ of Green functions, these
fields are coupled to external
sources $j_i$ $(i = 1,\ldots,n_B)$, $\rho_a$, $\ol \rho_a$
$(a = 1,\ldots,n_F)$,
\beqa
\label{funcint}
Z[j,\rho,\ol \rho] =
e^{iW[j,\rho,\ol \rho]} =
\int [d\vp d\psi d\ol \psi] \,
e^{i(S[\vp,\psi,\ol \psi] + j^T \vp + \ol \psi \rho + \ol \rho \psi)}~,
\eeqa
where $W[j,\rho,\ol \rho]$ is the generating functional of connected
Green functions. I have used the notation
\beq
\label{notation}
j^T \vp + \ol \psi \rho + \ol \rho \psi := \int d^dx ~ ( j_i \vp_i +
\ol \psi_a \rho_a + \ol \rho_a\psi_a)~. 
\eeq
The normalization of the functional integral is determined by
the condition $Z[0,0,0] = 1$.
The solutions of the classical equations of motion
\beq
\frac{\delta S}{\delta \vp_i} + j_i = 0, \quad
\frac{\delta S}{\delta \ol \psi_a} + \rho_a = 0, \quad
\frac{\delta S}{\delta \psi_a}  - \ol \rho_a = 0
\label{EOM}
\eeq
are denoted by $\vp_{\rm cl}$, $\psi_{\rm cl}$. They are uniquely
determined functionals of the external sources.
With fluctuation fields $\xi,\eta$ defined by
\beqa
\label{ff}
\vp_i &=& \vp_{{\rm cl},i} + \; \xi_i~, \no \\
\psi_a &=& \psi_{{\rm cl},a} + \eta_a ~,
\eeqa
the integrand in (\ref{funcint}) is expanded in terms of
$\xi,\eta,\ol \eta$. 
The resulting loop expansion of the generating functional
\bea
W = W_{L = 0} + W_{L = 1} + \ldots \no
\eea
starts with the classical action in the presence of external
sources:
\beq
W_{L = 0} = S[\vp_{\rm cl},\psi_{\rm cl},\ol \psi_{\rm cl}]
+ j^T \vp_{\rm cl} + \ol \psi_{\rm cl} \rho + \ol \rho \psi_{\rm cl}~.
\eeq
The one-loop term $W_{L = 1}$ is given by a Gaussian functional
integral
\beqa
\label{Gauss}
e^{iW_{L = 1}} =
\int [d\xi d\eta d\ol \eta] ~
e^{i S^{(2)}[\vp_{\rm cl},\psi_{\rm cl}, \ol \psi_{\rm cl};
\xi, \eta,\ol \eta]}~,
\eeqa
where
\beq
S^{(2)}[\vp_{\rm cl},\psi_{\rm cl}, \ol \psi_{\rm cl}; \xi, \eta,\ol \eta] =
\int d^dx ~ \cL^{(2)}(\vp_{\rm cl},\psi_{\rm cl}, \ol \psi_{\rm cl};
\xi, \eta,\ol \eta)
\label{S2}
\eeq
is quadratic in the fluctuation variables.
Employing the notation introduced in (\ref{notation}), $S^{(2)}$
takes the general form
\beqa
\label{Sfluc}
S^{(2)} &=& \dfrac{1}{2}\xi^TA\xi + \ol \eta B \eta + \xi^T \ol \Gamma \eta +
\ol \eta \Gamma \xi \nn
& = & \dfrac{1}{2}\left(\xi^T A\xi + \ol \eta B \eta - 
\eta^T B^T\ol\eta^T + \xi^T \ol \Gamma \eta - \eta^T \ol \Gamma^T \xi +
\ol \eta \Gamma \xi -  \xi^T \Gamma^T \ol\eta^T \right) ~,
\eeqa
where $A,B,\Gamma, \ol \Gamma$ are operators in the respective spaces;
$A = A^T$  and $B$ are bosonic differential operators, whereas $\Gamma$
and
$\ol \Gamma$ are fermionic (Grassmann) operators.  
They all depend on the classical solutions $\vp_{\rm cl}$, $\psi_{\rm cl}$.

The standard procedure for the evaluation of (\ref{Gauss}) is to
integrate first over the fermion fields
$\eta,\ol \eta$ to yield the bosonic functional integral
\beqa
e^{iW_{L = 1}} =
 \det B \int [d\xi] ~
e^{ \frac{i}{2}  \xi^T(A - \ol \Gamma B^{-1} \Gamma + 
\Gamma^T B^{-1T} \ol \Gamma^T)\xi}~.\nonumber
\eeqa
This leads to  the familiar result
\beqa
\label{W1}
W_{L = 1} &=& \dfrac{i}{2} \left[\ln \det (A - \ol \Gamma B^{-1} \Gamma + 
\Gamma^T B^{-1T} \ol \Gamma^T) -\ln \det A_0\right]
- i (\ln \det B - \ln \det B_0) \nn
&=& \dfrac{i}{2} \mbox{ Tr } \ln \dfrac{A}{A_0}
- i \mbox{ Tr } \ln \dfrac{B}{B_0} + \dfrac{i}{2}
\mbox{ Tr } \ln(1 - A^{-1} \ol \Gamma B^{-1} \Gamma 
+ A^{-1}\Gamma^T B^{-1T} \ol \Gamma^T) \no \\
&=& \dfrac{i}{2} \mbox{ Tr } \ln \dfrac{A}{A_0}
- i \mbox{ Tr } \ln \dfrac{B}{B_0} -
\sum_{n=1}^\infty \dfrac{i}{2n}\mbox{Tr}\left(A^{-1} \ol \Gamma B^{-1}
\Gamma
- A^{-1} \Gamma^T B^{-1T} \ol \Gamma^T \right)^n , \nn [4pt]
&& A_0 := A|_{j = \rho = \ol \rho = 0}, \quad
B _0 := B|_{j = \rho = \ol \rho = 0}~. 
\label{eqconv}
\eeqa
Recalling that $A^{-1}, B^{-1}$ are the scalar and fermion matrix
propagators in the presence of external sources, the one-loop functional
$W_{L = 1}$ is seen to be a sum of the bosonic one-loop functional
$\frac{i}{2} \mbox{ Tr }\ln (A/A_0)$, the fermion-loop functional
$-i \mbox{ Tr } \ln (B/B_0)$ and a mixed one-loop functional where scalar
and
fermion propagators alternate. In order to determine the
ultraviolet divergences that occur in the mixed term in
(\ref{eqconv}),
 the calculational inconveniences mentioned
in Sect. \ref{sec: Introduction} are encountered.

These problems can be circumvented \cite{NGE98} by  reorganizing
the three parts of $W_{L = 1}$ into a more compact form, using the notion
of
supermatrices, supertraces, etc. (see for instance \cite{ANZ,berezin}).
Combining the bosonic
and fermionic fluctuation variables in a multicomponent field
\beq
\lambda = \left[ \ba{c}
\xi \\ 
\eta \\
\ol \eta^T \ea
\right]~ ,
\eeq
$S^{(2)}$ in (\ref{Sfluc}) can be written as
\beqa
S^{(2)} = \dfrac{1}{2} \lambda^T\; K \; \lambda~. 
\eeqa
The explicit form of the supermatrix operator $K$ follows immediately from
the second line in (\ref{Sfluc}): 
\beq
K = \left[ \ba{ccc}
A & \ol \Gamma & -\Gamma^T \\
- \ol \Gamma^T & 0 & - B^T \\
\Gamma & B & 0
\ea \right]~.
 \label{KSUSY}
\eeq
The one-loop functional of connected Green functions can now be written in
compact form  \cite{NGE98} in terms of a supertrace 
\beq
W_{L = 1} = \frac{i}{2} \mbox{ Str } \ln \frac{K}{K_0}~. 
\label{Zsuper}
\eeq
With the notation
\bea
\mbox{Str } O = \int d^dx ~ \mbox{str} \langle x|O|x \rangle \no
\eea
I distinguish supertraces with and without space-time integration. 

For actual calculations, the form of the supermatrix operator $K$ defined
in (\ref{KSUSY}) is not the most convenient one. Applying a similarity
transformation to $K$, the generating functional can also be written as
\beq
W_{L = 1} = \frac{i}{2} \mbox{ Str } \ln \frac{K'}{K'_0} 
\label{Zdiag}
\eeq
with
 \beq
K' = \left[ \ba{ccc}
A & \sqrt{\mu} ~ \ol \Gamma & -  \sqrt{\mu} ~ \Gamma^T \\
\sqrt{\mu} ~  \Gamma & \mu B & 0  \\
\sqrt{\mu} ~ \ol \Gamma^T & 0 & \mu B^T
\ea \right]~.
 \label{KP}
\eeq 
The arbitrary  mass parameter $\mu$ introduced in (\ref{KP})  
guarantees equal dimensions for all entries
in $K'$ ($[K']=[A]=2$). Although this quantity does not, of course, appear
in any final result,  it turns out to be quite helpful for the inspection
of expressions at intermediate stages of calculations.

In the proper-time formulation, the one-loop functional assumes the
form 
\beqa
W_{L = 1} &=& - \dfrac{i}{2} \int_0^\infty \dfrac{d\tau}{\tau} \mbox{ Str
} 
\left(e^{i\tau K'}-e^{i\tau K'_0}\right) \nn
&=&  - \dfrac{i}{2} \int_0^\infty \dfrac{d\tau}{\tau} \int d^dx 
\mbox{ str } 
\langle x|e^{i\tau K'}-e^{i\tau K'_0}|x \rangle ~, 
\label{ptf}
\eeqa
which is just the desired super-heat-kernel representation. Note that the 
convergence of the integral at the upper end ($\tau \ra \infty$) is
guaranteed by the small imaginary parts present in the bosonic and
fermionic differential operators $A$ and $B$, which are ensuring at the
same time the usual Feynman boundary conditions. (For a free theory $A = 
- \Box - M^2 + i \epsilon$, $B = i\not\!\partial - m + i \epsilon$.) On
the
other hand, the
behaviour of the integral at the lower end exhibits the divergence
structure of the theory under investigation.    

As long as we are only interested in those parts of the one-loop
functional that are at most bilinear in fermion fields, 
the supermatrix $K'$ can be reduced to the simpler form
 \beq
K'' = \left[ \ba{cc}
A & \sqrt{2\mu}~\ol \Gamma  \\
\sqrt{2\mu} ~ \Gamma & \mu B 
\ea \right] \scs
 \label{KPP}
\eeq
such that the one-loop functional reads
\beq
W_{L=1} = \dfrac{i}{2} \mbox{ Str } \ln \dfrac{K''}{K''_0}
- \dfrac{i}{2} \mbox{ Tr } \ln \dfrac{B}{B_0}
 + \ldots \label{Zss}
\eeq
The terms omitted are at least quartic in the fermion fields.

\section{Scalars Interacting with Heavy Fermions}
\label{sec: SFI}
\renewcommand{\theequation}{\arabic{section}.\arabic{equation}}
\setcounter{equation}{0}

In the case of chiral perturbation theory with heavy baryons, the
fluctuation action (\ref{Sfluc}) generated by the
lowest order meson--baryon Lagrangian ($\cO(p^2)$ in the mesonic and
$\cO(p)$ in the
baryonic part) has the general form
\beq
\label{HBfluc}
S^{(2)} = -\dfrac{1}{2}\xi^T(D_{\mu}D^{\mu} + Y)\xi +
\ol \eta (\alpha + \beta_{\mu} D^{\mu}) \xi  
 + \xi^T (\ol \delta -\ol \beta_{\mu} \D^{\mu}) \eta 
+ \ol \eta i v_{\mu} \D^{\mu} \eta  ~,
\eeq
where 
\beq
\ba{ll}
D_\mu = \del_\mu + X_\mu~, & \D_\mu = \del_\mu + f_\mu~, \\[4pt]
\ol \delta = \ol \alpha - \ol {\D_\mu \beta^\mu}~,     \\
v^2 = 1~, & v \cdot \beta = 0~. \ea
\label{explic}
\eeq
$X_\mu$, $Y$ and $f_\mu$ are bosonic (matrix-) fields, 
whereas $\alpha$ and $\beta_\mu$ are fermionic objects. The form of $\ol
\delta$ in (\ref{explic}) is required by the reality of (\ref{HBfluc}).
$v_\mu$ is the usual velocity vector introduced in 
the heavy mass expansion. Apart from the condition $v \cdot \beta = 0$
(which is indeed satisfied in heavy baryon chiral perturbation theory),
no further assumption about the terms entering in (\ref{HBfluc}) is
made in this section. The 
discussion will therefore  apply to a rather large class of theories
of scalars interacting with heavy fermions, not necessarily related to 
chiral perturbation theory. The further specialization to the
pion--nucleon 
system is reserved until the next chapter.

The action (\ref{HBfluc}) is invariant under local gauge transformations
\beq
\ba{llll}
\xi(x) &\ra& R(x) \xi(x)~, & R(x)^T R(x) =  {\bf 1}~,  \\
\eta(x) &\ra& U(x) \eta(x)~, & U(x)^{\dagger} U(x) = {\bf 1}~,  \\
X_\mu &\ra& R \del_\mu R^{-1} + R X_\mu R^{-1}~, & \\
Y &\ra& R Y R^{-1}~, & \\
f_\mu &\ra& U \del_\mu U^{-1} + U f_\mu U^{-1}~, & \\
\alpha &\ra& U \alpha R^{-1}~, & \\
\beta_\mu &\ra& U \beta_\mu R^{-1}~. & 
\ea
\label{gaugetransf}
\eeq
Consequently, also the divergent part of the one-loop functional exhibits 
this symmetry property \cite{thooft}. The matrix-fields $Y$, $\alpha$,
$\beta_\mu$
together with their covariant derivatives
\beqa
{\wh \nabla}_\mu Y &:=& \del_\mu Y + [X_\mu,Y], \no \\ 
{\wh \nabla}_\mu \alpha &:=& \del_\mu \alpha + f_\mu \alpha - \alpha
X_\mu, \no
\\
{\wh \nabla}_\mu \beta_\nu &:=& \del_\mu \beta_\nu 
+ f_\mu \beta_\nu - \beta_\nu X_\mu, 
\label{covdev} 
\eeqa
and the associated ``field-strength'' tensors
\beqa
X_{\mu \nu} &:=& \del_\mu X_\nu - \del_\nu X_\mu + [X_\mu,X_\nu], \no \\
f_{\mu \nu} &:=& \del_\mu f_\nu - \del_\nu f_\mu + [f_\mu,f_\nu] 
\label{fieldstrengths}
\eeqa
are therefore the appropriate building blocks for the construction 
of a gauge-invariant action.

The general heat-kernel formalism of the preceding section will now be applied
to (\ref{HBfluc}). In this case, the matrix-operators $A$, $B$, $\Gamma$ 
and $\ol \Gamma$ 
defined in (\ref{Sfluc}) are given by 
\beq
A = -D^2 - Y~, \quad
B = i v \cdot \D~, \quad 
\Gamma = \alpha + \beta \cdot D~,  \quad
\ol \Gamma = \ol \delta - \ol \beta \cdot \D~.
\eeq
As I am considering only terms at most bilinear in the fermionic
variables, the form (\ref{KPP}) for the supermatrix operator is
the appropriate one.
Employing the
method of Ball \cite{Ball}, the relevant diagonal space-time matrix element 
can be written as
\beqa
\mbox{str } \langle x|e^{i\tau K''}|x \rangle &=& \mbox{str} 
\int d^dk \langle x|e^{i\tau
K''}|k \rangle \langle k|x \rangle =
\mbox{str} \int \dfrac{d^dk}{(2\pi)^d}e^{ikx} e^{i\tau K''} e^{-ikx} \nn 
 &=&  \mbox{str} \int \dfrac{d^dk}{(2\pi)^d} e^{i\tau {\wt K}''} 
{\bf 1}~, 
\label{matrel}
\eeqa
with
\beq
{\wt K}'' = \left[ \ba{cc}
-D^2-Y+k^2+2ik \cdot D & \sqrt{2 \mu} (\ol \delta - \ol \beta \cdot \D 
+ i k \cdot \ol \beta)  \\
\sqrt{2 \mu} ( \alpha + \beta \cdot D - i k \cdot \beta) 
& \mu (iv \cdot \D + v \cdot k)
\ea \right]~. 
\eeq
The further evaluation  of this expression is considerably simplified by the 
observation
that in the following intermediate steps we may restrict ourselves to constant 
fields \cite{Ball} $X_\mu$, $\alpha$, $\beta_\mu$, $f_\mu$, $Y = -X^2$. As
the 
final result for the one-loop divergences has to be gauge-invariant, no 
information 
is lost and the full expression for space-time dependent fields is recovered 
by the  substitutions
\beqa
-X^2 &\ra& Y~, \nn
-[X_\mu,X^2] &\ra& {\wh \nabla}_\mu Y~, \nn
\left [ X_\mu,X_\nu \right ] &\ra& X_{\mu \nu}~, \nn
f_\mu \alpha - \alpha X_\mu &\ra& {\wh \nabla}_\mu \alpha~, \nn
f_\mu \beta_\nu - \beta_\nu X_\mu &\ra& {\wh \nabla}_\mu \beta_\nu~, \nn
\left [ f_\mu,f_\nu \right ] &\ra& f_{\mu \nu}~.
\label{substitutions}
\eeqa
In this approach, (\ref{matrel}) reduces to the much
simpler expression
\beq
\label{exponential}
\mbox{str } \langle x|e^{i\tau K''}|x \rangle =  
\int \dfrac{d^dk}{(2\pi)^d} \, \mbox{ str }  e^{M+N}
\eeq
with
\beqa
M &=& i \tau \left[ \ba{cc}
k^2 + 2 i k\cdot X & 0 \\
0 & \mu (i v \cdot f + v \cdot k) \ea \right]~, \nn [4pt]
N &=& i \tau \sqrt{2 \mu} \left[ \ba{cc}
0 & \ol \delta - \ol \beta \cdot f + i k \cdot \ol \beta  \\
 \alpha + \beta \cdot X  - i k \cdot \beta & 0
\ea \right] ~. \label{MN}
\eeqa

Let us first consider the  part bilinear in the fermionic
matrix $N$ (generating the terms of the form  $\ol \alpha \ldots
\alpha$, $\ol \alpha \ldots \beta_\mu$, $\ol \beta_\mu \ldots \alpha$,
$\ol \beta_\mu \ldots \beta_\nu$). The corresponding part of the
generating functional (\ref{W1}) is just
\beq
W_{L=1}|_{\ol \Gamma \ldots \Gamma} := - i \mbox{ Tr } (A^{-1} \ol \Gamma
B^{-1} \Gamma)~.
\eeq
The appropriate decomposition of the exponential in (\ref{exponential})
can be performed by using  Feynman's ``disentangling'' theorem \cite{Feynman}:
\bea
\label{Duhamel}
\mbox{exp}(M+N) = \mbox{exp} \, M ~ \mbox{P}\!_s \, \mbox{exp} 
\int_0^1 ds \, \wt{N} (s)                 
\eea
with
\bea
\wt{N} (s) := e^{-sM} N e^{sM} \no
\eea
and
\bea
\mbox{P}\!_s \, \mbox{exp} \int_0^1 ds \, \wt{N} (s) :=
\sum_{n=0}^{\infty} \int_0^1 ds_1 \int_0^{s_1} ds_2 \ldots \int_0^{s_{n-1}}
ds_n \, \wt{N} (s_1) \wt{N} (s_2) \ldots \wt{N} (s_n)~.
\eea
(In the mathematical literature, (\ref{Duhamel}) is also known as
``Duhamel's formula''.) Picking out the part bilinear in
$N$,
\beq
\mbox{ str } e^{M+N} = \int_0^1 ds \int_0^s ds'
\mbox{ str }\left[ e^{(1-s)M}N e^{(s-s')M}N e^{s'M} \right] +
\ldots~, 
\label{dtheorem}
\eeq
 a few simple manipulations lead to 
\beqa
\mbox{ str }  e^{M+N} 
&=& - 2 \mu \tau^2   \int_0^1 dz \,
e^{i \tau [(1-z)k^2+z\mu v \cdot k]}  
 \mbox{ tr } \left[ (\ol \delta - \ol \beta \cdot f + i \ol \beta \cdot
k)
e^{-\tau z \mu v \cdot f} \right. \nn
&& \qquad \left. (\alpha + \beta \cdot X - i \beta \cdot k)
e^{-2\tau (1-z) k \cdot X}\right] + \ldots 
\eeqa
After integration over $z$, the $\mu$-dependent terms cancel once the
proper-time and the momentum-space integrals are applied. The
remaining contribution to $W_{L=1}$ assumes the form
\beqa
W_{L=1}|_{\ol \Gamma \ldots \Gamma}
&=& - \int d^dx \int_0^\infty \dfrac{dt}{t} 
t^{3-d} \int \dfrac{d^dl}{(2\pi)^d} \, e^{iv \cdot l}  \mbox{ tr }
\left[(\ol \delta-\ol \beta \cdot f + i \ol \beta \cdot l /t)
e^{-t v \cdot f} \right. \nn
&& \qquad \left. (\alpha + \beta \cdot X - i \beta \cdot l /t)
(l^2+2 i t l \cdot X)^{-1}\right]~,
\eeqa
where a suitable change of the integration variables has been performed.
The divergent part (for $d \ra 4$) can now be easily isolated:
\beqa
W_{L=1}^{\rm div}|_{\ol \Gamma \ldots \Gamma} &=&
\Gamma(4-d) \int d^dx
\int \dfrac{d^dl}{(2\pi)^d} \, \dfrac{e^{i v \cdot l}}{l^2} \nn
&& 
\mbox{ tr } \left \{ (\ol \delta - \ol \beta \cdot f) \, v \cdot
f \, 
(\alpha +
\beta \cdot X) 
+\dfrac{1}{3!} \ol \beta \cdot l \, (v \cdot f)^3 \beta \cdot l \right. 
\nn
&& \qquad  \left. + \left  [ (\ol \delta - \ol \beta \cdot f)
(\alpha + \beta \cdot X) 
+i(\ol \delta - \ol \beta \cdot f) \, v \cdot f \, \beta \cdot l
\right. \right. 
\nn
&& \qquad \left. \left.  -i \ol \beta \cdot l \, v \cdot f \, 
(\alpha + \beta \cdot X)
+\dfrac{1}{2!} \ol \beta \cdot l \, (v \cdot f)^2
\beta
\cdot l \right ]
\dfrac{2 i l \cdot X}{l^2} \right. \nn 
&& \qquad \left. +\left[-i (\ol \delta - \ol \beta \cdot f) \, \beta \cdot
l +i\ol \beta \cdot l \, (\alpha + \beta \cdot X) 
-\ol \beta \cdot l \, v \cdot f \, \beta \cdot l \right]
\dfrac{4 (l \cdot X)^2}{(l^2)^2} \right.  \nn
&& \qquad  \left. - \ol \beta \cdot l \, \beta \cdot l \, \dfrac{8 i (l
\cdot X)^3}{(l^2)^3} \right \}~.
\label{divergences}
\eeqa
The necessary formulas for the $l$-integration are
given in the Appendix. In the last step, one has to identify the
appropriate
gauge-invariant combinations (constituting a non-trivial check of the
calculation) and reconstruct the full result by using (\ref{substitutions}).
In this way, I finally obtain:
\beqa
W_{L=1}^{\rm div}|_{\ol \Gamma \ldots \Gamma} &=& 
\dfrac{i}{48 {\pi}^2 (d-4)} \int d^4x \mbox{ tr }
\left \{ - 12 \, \ol \alpha \, v \cdot {\wh \nabla} \alpha 
 + 6 \left [ \ol \alpha \beta_\mu X^{\mu \nu} v_\nu
         +\ol \beta_\mu \alpha X^{\mu \nu} v_\nu \right ] \right. \nn
&&\left. - 3 \left [ \ol \beta \cdot \beta \, v \cdot {\wh \nabla} Y
         + 2 \, \ol \beta_\mu (v \cdot {\wh \nabla} \beta^{\mu}) Y \right
] 
- 4 \ol \beta_\mu (v \cdot {\wh \nabla})^3 \beta^{\mu}
+ \ol \beta \cdot \beta \, {\wh \nabla}_\mu X^{\mu \nu}
v_\nu \right. \nn
&&\left.  + 6 \, \ol \beta_\mu (v \cdot {\wh \nabla} \beta_\nu)
X^{\mu \nu}
   + 4 \, \ol \beta _\mu \beta_\nu \, v \cdot {\wh \nabla} X^{\mu
\nu}
   + 2 \, \ol \beta_\mu \beta_\nu {\wh \nabla}^{\mu} X^{\nu \rho}
      v_{\rho} \right \}~. 
\label{result}
\eeqa
Note that (\ref{result}) has to be real, which is another independent
check of the result.

The remaining part of the generating functional with the fermionic
operators $\Gamma$, $\ol \Gamma$ turned off, 
\beq
W_{L=1}|_{\Gamma = \ol \Gamma = 0} =
\dfrac{i}{2} \mbox{ Tr } \ln \dfrac{A}{A_0}
- i \mbox{ Tr } \ln \dfrac{B}{B_0}~,
\label{W10}
\eeq
does not require any additional effort. A simple calculation (involving a
Gaussian momentum-space integration) gives
\beq
\label{Wstandard}
\dfrac{i}{2} \mbox{ Tr } \ln A |_{\rm div} =
- \dfrac{1}{(4\pi)^2 (d-4)} \int d^4x \mbox{ tr} \left( \dfrac{1}{12}
X_{\mu \nu} X^{\mu \nu} + \dfrac{1}{2} Y^2 \right)~,
\eeq
which is the standard result obtained by 't Hooft \cite{thooft} using
diagrammatic methods.

The second term in (\ref{W10}) vanishes identically, as it corresponds to
the closed loop of a ``light'' fermion component in the heavy mass
expansion:
\beqa
 \mbox{Tr } \ln B &=& - \int d^dx \int_0^\infty \dfrac{dt}{t} 
 \int \dfrac{d^dk}{(2\pi)^d} \mbox{ tr } \left( 
e^{it(iv \cdot \D + v \cdot k)} {\bf 1} \right) \nn
&=&  - \int d^dx \int_0^\infty \dfrac{dt}{t} t^{-d}
 \int \dfrac{d^dl}{(2\pi)^d} e^{iv \cdot l}
\mbox{ tr } \left( 
e^{- t v \cdot \D} {\bf 1} \right) = 0~, \no
\eeqa
which follows from
\beqa
\int \dfrac{d^dl}{(2\pi)^d} e^{iv \cdot l} = \delta^{(d)}(v) = 0~. \no
\eeqa

\section{Renormalization of the Pion--Nucleon Interaction}
\label{sec: renorm}
\renewcommand{\theequation}{\arabic{section}.\arabic{equation}}
\setcounter{equation}{0}
The functionals (\ref{result}) and(\ref{Wstandard}) are the basic formulas
for the analysis of the one-loop divergences to $\cO(p^3)$ in heavy baryon
chiral perturbation theory. They can be applied to both the two-flavour
and the
three-flavour case. In the following I shall confine myself to  chiral
$SU(2)$. 

The starting point for the formulation of the effective field theory of
the pion--nucleon system is QCD  with the two light flavours $u,d$
coupled to external Hermitian fields \cite{GLAP}:
\beq
\cL = \cL^0_{\rm QCD} + \bar q \gamma^\mu \left(\V_\mu + \frac{1}{3}
\V^s_\mu + \gamma_5 \A_\mu\right) q - \bar q (S - i\gamma_5 P)q~, \qquad
q = \left[ \ba{c} u \\ d \ea \right]~. \label{QCD}
\eeq
$\cL^0_{\rm QCD}$ is the QCD Lagrangian with $m_u = m_d = 0$,
$S$ and $P$ are general two-dimensional matrix fields, the isotriplet
vector
and axial-vector fields $\V_\mu,\A_\mu$ are traceless and the
isosinglet vector field $\V^s_\mu$ is included to generate the
electromagnetic current. 

Explicit chiral symmetry breaking is built in by setting
$S = \M_{\rm quark} = \mbox{diag }[m_u,m_d]$.
The chiral group $G = SU(2)_L \times SU(2)_R$ is spontaneously broken to
the isospin group
$SU(2)_V$. It is realized non-linearly \cite{CCWZ}
on the Goldstone pion fields $\phi$:
\beqa
u_L(\phi) & \stackrel{g}{\ra} & g_L u_L(\phi) h(g,\phi)^{-1}, \qquad
g = (g_L,g_R) \in G~, \no \\
u_R(\phi) & \stackrel{g}{\ra} & g_R u_R(\phi) h(g,\phi)^{-1},
\label{coset}
\eeqa
where $u_L, u_R$ are elements of the chiral coset space
$SU(2)_L \times SU(2)_R/SU(2)_V$ and the compensator field
$h(g,\phi)$ is in $SU(2)_V$.

The nucleon doublet $\Psi$ transforms as
\beq
\Psi = \left[ \ba{c} p \\ n \ea \right] \stackrel{g}{\ra}
\Psi' = h(g,\phi)\Psi \label{psi}
\eeq
under chiral transformations. The local nature of this transformation
requires a connection
\beq
\Gamma_\mu = \frac{1}{2} \left [ u_R^\dg (\partial_\mu - i r_\mu)u_R +
u_L^\dg (\partial_\mu - i \ell_\mu)u_L\right ] \label{conn}
\eeq
in the presence of external gauge fields
\beq
r_\mu = \V_\mu + \A_\mu~, \qquad \ell_\mu = \V_\mu - \A_\mu
\label{gf}
\eeq
to define a covariant derivative
\beq
\nabla_\mu \Psi = (\partial_\mu + \Gamma_\mu - i \V^s_\mu)\Psi~.
\eeq

To lowest order in the chiral expansion the effective Lagrangian
of the pion--nucleon system is \cite{GSS88,GLAP}
\beqa
\cL_{\rm eff} &=& \frac{F^2}{4} \langle u_\mu u^\mu + \chi_+ \rangle
+ \bar \Psi (i \not\!\nabla - m + \frac{g_A}{2}
\not\!u \gamma_5) \Psi~,
\label{Leff} 
\eeqa
with
\beqa
u_\mu &=& i \left [ u^{\dagger}_R (\del_\mu - i r_\mu) u_R
                   - u^{\dagger}_L (\del_\mu - i \ell_\mu) u_L \right ]~,
\nn
\chi &=& 2B(S+iP), \qquad \chi_{+} = u^{\dagger}_R \chi u_L
+ u^{\dagger}_L \chi^{\dagger} u_R~. \no
\eeqa
$F,m,g_A$ are the pion decay constant, the nucleon mass and the neutron
decay constant in the
chiral limit, whereas $B$ is related to the quark condensate.  $\langle
\dots \rangle$ stands for the trace in
flavour space.

The heavy baryon mass expansion of (\ref{Leff}) is obtained by introducing
velocity-dependent fields
\beqa
N_v(x) &=& e^{i m v \cdot x} P_v^+ \Psi(x)~, \label{vdf} \\
H_v(x) &=& e^{i m v \cdot x} P_v^- \Psi(x)~, \no \\
P_v^\pm &=& \frac{1}{2} (1 \pm \not\!v)~, \qquad v^2 = 1 ~, \no
\eeqa
leading to
\beqa
\cL_{\rm eff} &=& \frac{F^2}{4} \langle u_\mu u^\mu + \chi_+ \rangle 
+ \ol N_v (  iv \cdot \nabla + g_A S \cdot u ) N_v + \ldots
\label{LHB} 
\eeqa
The additional terms involving the ``heavy'' fermion components $H_v$ are
irrelevant for our present purposes. (For a more detailed discussion the
reader is referred to \cite{Eck94,EM95}.) The only dependence on
Dirac matrices in (\ref{LHB}) is through the spin-vector matrices
\beq
S^\mu = \frac{i}{2} \gamma_5 \sigma^{\mu\nu} v_\nu~, \qquad
S \cdot v = 0~, \qquad S^2 = - \frac{3}{4} {\bf 1}~,
\eeq
which obey the (anti-) commutation relations
\beq
\{ S^\mu,S^\nu\} = \frac{1}{2} (v^\mu v^\nu - g^{\mu\nu})~, \qquad
[S^\mu,S^\nu] = i \ve^{\mu\nu\rho\sigma} v_\rho S_\sigma~.
\eeq

To obtain the associated second-order fluctuation Lagrangian
$\cL_{\rm eff}^{(2)}$, (\ref{LHB}) is expanded around the classical fields
$\phi_{\rm cl}$, $N_{v, \rm cl}$. In the standard ``gauge'' $u_R(\phi_{\rm
cl}) = u_L^{\dagger}(\phi_{\rm cl}) =: u(\phi_{\rm cl})$ a convenient
choice of the bosonic
fluctuation variables $\xi_i \, (i=1,2,3)$ is given by \cite{GLAP}
\beq
u_R(\phi) = u(\phi_{\rm cl}) \,
e^{\frac{i\stackrel{\ra}{\xi}\!{(\phi)}
\cdot \stackrel{\ra}{\tau}}{2
F}}, \qquad 
u_L(\phi) = u^\dg(\phi_{\rm cl}) \,
e^{-\frac{i\stackrel{\ra}{\xi}\!{(\phi)}
\cdot \stackrel{\ra}{\tau}}{2
F}}, \qquad
\stackrel{\ra}{\xi}\!{(\phi_{\rm cl})} = 0~,
\eeq
where $\stackrel{\ra}{\tau}$ denotes the Pauli matrices. For the fermion
fields I write $N_v = N_{v, {\rm cl}} + \eta$.
In this way I get
\beqa
\cL_{\rm eff}^{(2)} &=& \dfrac{1}{2} \left[ (d_{\mu \, ki} \xi_i)
(d^{\mu}_{kj} \xi_j) - \sigma_{ij} \xi_i \xi_j \right] \nn
&& + \dfrac{1}{8 F^2} {\ol N}_v \mbox{\Large [} i \xi_i [\tau_i,\tau_k] (v
\cdot
d_{kj} \xi_j) + g_A \xi_i [\tau_i,[S \cdot u,\tau_j]] \xi_j \mbox{\Large
]} N_v \nn
&& + \dfrac{1}{F} {\ol N}_v \left [ \dfrac{i}{4} [v \cdot u, \tau_i] \xi_i
-g_A S_\mu \tau_i (d^{\mu}_{ij} \xi_j) \right ] \eta \nn
&& + \dfrac{1}{F} \ol \eta  \left [ \dfrac{i}{4} [v \cdot u, \tau_i] \xi_i
-g_A S_\mu \tau_i (d^{\mu}_{ij} \xi_j) \right ] N_v \nn
&& + \ol \eta (i v \cdot \nabla + g_A S \cdot u) \eta~,
\label{Leff2}
\eeqa
where
\beqa
d^\mu_{ij} &=& \delta_{ij} \partial^\mu + \gamma^\mu_{ij}~, \qquad
\gamma^\mu_{ij} = - \frac{1}{2} \langle \Gamma^\mu
[\tau_i,\tau_j]\rangle~, \nn
\sigma_{ij} &=& \dfrac{1}{4} \langle (u \cdot u + \chi_+) \delta_{ij}
- \tau_i u_\mu \tau_j u^{\mu} \rangle~.
\eeqa
Note that the quantities $N_v$, $u_\mu$, $\Gamma_\mu$, $\chi_+$ in
(\ref{Leff2}) are to be taken at the solutions of the classical equations
of motion. (The subscript ``cl'' has only been dropped for simplicity.) 

It is now easy to verify that the action associated with (\ref{Leff2}) can
indeed be written in the standard form (\ref{HBfluc}) by setting
\beqa
X_\mu &=& \gamma_\mu + g_\mu~, \qquad g^{\mu}_{ij} = - \dfrac{i v^{\mu}}{8
F^2} {\ol N}_v [\tau_i,\tau_j] N_v~, \qquad i=1,2,3~,\nn
Y &=& \sigma + s~, \qquad \quad s_{ij} = \dfrac{g_A}{4 F^2} {\ol N}_v
\left (2 
\,
\delta_{ij} \, S \cdot u -\tau_i \, S \cdot u \, \tau_j - \tau_j \, S
\cdot u \, \tau_i
\right ) N_v~, \nn
f_\mu &=& \Gamma_\mu - i \V^s_\mu - i v_\mu g_A S\cdot u \nn
\alpha_{ai} &=& \dfrac{i}{4F} \left([v \cdot u, \tau_i] N_v \right )_a~,
\qquad a=1,2~, \nn
(\beta_\mu)_{ai} &=& - \dfrac{g_A}{F} S_\mu (\tau_i N_v)_a~.
\label{insert}
\eeqa 

Let us first consider the one-loop divergences generated by
(\ref{Wstandard}). Using (\ref{insert}),
\beqa
\mbox{tr } Y^2 &=& \mbox{tr } \sigma^2 + 2 \mbox{ tr }(\sigma s) + \dots
\label{Y*Y}
\eeqa 
and
\beqa
\mbox{tr } (X_{\mu \nu} X^{\mu \nu})&=& \mbox{tr } \gamma_{\mu \nu}
\gamma^{\mu \nu} + 4\mbox{ tr }
\left \{ \gamma_{\mu \nu} (\del^{\mu} g^{\nu} +[\gamma^{\mu},g^{\nu}])
\right \} + \ldots ~, 
\label{X*X}
\eeqa
where
\beq
\gamma_{\mu \nu} := \del_\mu \gamma_\nu - \del_\nu \gamma_\mu +
[\gamma_\mu,\gamma_\nu]~.
\eeq
The first terms on the right-hand sides of (\ref{Y*Y}) and (\ref{X*X})
are purely mesonic; they determine the divergence structure of the
well-known  Gasser--Leutwyler
functional of $\cO(p^4)$ \cite{GLAP}. The second ones are bilinear in the
fermion
fields, whereas the dots refer to irrelevant terms $\sim (\ol N_v \dots
N_v)^2$. To facilitate the comparison with \cite{Eck94}, I write
the fermion bilinears extracted from (\ref{Wstandard}) in the
following form:
\beqa
W_{L=1}^{\rm div}|_{\Gamma = \ol \Gamma = 0} = \int d^4x \, \ol N_v \,
\Sigma_1^{\rm div} \, N_v~, \qquad
\Sigma_1^{\rm div} &=& -\frac{1}{8 \pi^2 F^2 (d-4)} \wh \Sigma_1~.
\label{Si1div}
\eeqa
For $\wh \Sigma_1$ I find
\beqa
\wh \Sigma_1 &=&  - \frac{i}{6} \left( \nabla^\mu \Gamma_{\mu\nu} v^\nu
\right)
 + \frac{g_A}{4} \left( \langle u \cdot u + \chi_+ \rangle  S \cdot u 
+ \langle S \cdot u  \, u_\mu \rangle u^{\mu} \right ) \label{Sigma1}~,
\eeqa
where
\beq
\Gamma_{\mu\nu} := \partial_\mu \Gamma_\nu - \partial_\nu \Gamma_\mu +
[\Gamma_\mu,\Gamma_\nu]~. 
\eeq
This result agrees with the corresponding expression in (36) of
\cite{Eck94}. Note that I have used several $SU(2)$ relations  to
arrive at a simpler form for $\wh \Sigma_1$ in (\ref{Sigma1}).

The one-loop divergences originating from (\ref{result}) are again
presented in the form
\beqa
W_{L=1}^{\rm div}|_{\Gamma \ldots \ol \Gamma} = \int d^4x \, \ol N_v \,
\Sigma_2^{\rm div} \, N_v~,
\qquad
\Sigma_2^{\rm div} = -\frac{1}{8 \pi^2 F^2 (d-4)} \wh \Sigma_2~.
\label{Si2div}
\eeqa 
Inserting (\ref{insert}) in (\ref{result}), I obtain: 
\beqa
\wh \Sigma_2 &=& i\left\{\frac{1}{4} [2(v \cdot u)^2 +
\langle (v \cdot u)^2\rangle] v \cdot \nabla +
\frac{1}{2} v \cdot u(v \cdot \nabla v \cdot u) +
\frac{1}{4} \langle v \cdot u (v \cdot \nabla v \cdot u)\rangle\right\}
\no \\
&& \mbox{} + g_A \left\{- \frac{1}{2} v \cdot u \langle S \cdot u \, v
\cdot u\rangle + \frac{1}{4} \langle S \cdot u(v \cdot u)^2\rangle
- S^\mu v^\nu [\Gamma_{\mu\nu},v \cdot u] \right\} \no \\
&& \mbox{} + ig^2_A \left\{- \frac{3}{2} (v \cdot \nabla)^3 -
\frac{5}{6} \left( \nabla^\mu \Gamma_{\mu\nu} v^\nu \right) + i
\ve^{\mu\nu\rho\sigma}
v_\rho S_\sigma [2\Gamma_{\mu\nu} v \cdot \nabla +
(v \cdot \nabla \Gamma_{\mu\nu})] \right. \no \\
&& \mbox{} \left. - \frac{3}{32} \left(v \cdot \del \langle 4 u \cdot u 
+ 3 \chi_+\rangle \right)
-  \frac{3}{16}  \langle 4 u \cdot u 
+ 3 \chi_+\rangle  v \cdot \nabla\right\}
\no \\
&& \mbox{} + g^3_A \left\{- \frac{1}{2} S \cdot u(v \cdot \nabla)^2 -
2 S^\mu \langle S \cdot u \, \Gamma_{\mu\nu} \rangle S^\nu -
\frac{1}{2} (v \cdot \nabla S \cdot u) v \cdot \nabla \right. \no \\
&& \mbox{} \left. - \frac{1}{6} \left( (v \cdot \nabla)^2 S \cdot u
\right) -
\frac{1}{4} u_\mu \langle u^\mu S \cdot u \rangle - \frac{1}{16}
S \cdot u \langle \chi_+\rangle \right\} \no \\
&& \mbox{} + ig^4_A S_\mu \left\{ [ 2(S \cdot u)^2 - 4 \langle
(S\cdot u)^2\rangle ] v \cdot \nabla + \frac{2}{3}(v \cdot \nabla
S \cdot u) S \cdot u  \right.
\no \\
&& \mbox{} + \left. \frac{4}{3} S \cdot u (v \cdot \nabla S \cdot u)
-  4 \langle S \cdot u(v \cdot \nabla S \cdot u)\rangle
\right\} S^\mu \no \\
&& \mbox{} + g^5_A S_\mu \left\{ \frac{2}{3} (S \cdot u)^3
- \frac{4}{3} \langle (S \cdot u)^3\rangle \right\} S^\mu~,  \label{S2h}
\eeqa
which is in agreement with the corresponding result in
\cite{Eck94}. (Note that ``$+-$'' in the fourth line of (53) in
\cite{Eck94} should be read as a minus sign.)

\section{Conclusions}
\label{sec: Conclusions}
\renewcommand{\theequation}{\arabic{section}.\arabic{equation}}
\setcounter{equation}{0}
I have shown that the super-heat-kernel technique constitutes the
appropriate theoretical tool for analyzing the one-loop divergences in
systems with (non-renormalizable) boson--fermion interactions. 
I recall here the essential ingredients that were combined
to arrive at an efficient computational scheme:
\begin{itemize}
\item
The one-loop functional is written in terms of the superdeterminant of
a suitably chosen supermatrix operator.
\item 
The associated super-heat-kernel representation is the 
appropriate form of the one-loop functional for studying
its divergence structure.
\item
It is easier to determine the diagonal heat-kernel matrix elements
directly by inserting a complete set of plane waves instead of calculating 
the Seeley--DeWitt coefficients with two different space-time arguments
and taking the coincidence-limit at the end. 
\item 
The heat-kernel-representation is
perfectly well defined also for supermatrices with
first-order (fermion) differential operators\footnote{Note, however, that
``squaring'' of the fermionic differential operator may simplify the
analysis in theories where the full relativistic Dirac
operator is still present \cite{NGE98}.}.
\item
The second-order fluctuation action is invariant under a local gauge
transformation. As a consequence, this
symmetry property is also shared by the divergence functional.
\item
At intermediate stages, the calculation can be carried out with constant
(classical) fields, avoiding cumbersome manipulations with derivatives
acting on space-time dependent objects. At the end, the general result is
recovered by gauge invariance.
\item
Feynman's disentangling theorem allows the proper decomposition of the
exponential of a sum of non-commuting terms.
\item
With the divergence functional given in compact form, the one-loop
renormalization of effective quantum field
theories becomes an easy task, requiring only a few purely algebraic
operations.
\end{itemize}
The application to heavy baryon chiral perturbation theory with two
flavours served as an explicit example. A previous result for the
counterterms to $\cO(p^3)$ was confirmed.

With the super-heat-kernel method at hand, the systematic study of
effective field theories at the one-loop level is simplified considerably.
I am
giving here a small selection of possible applications and extensions of
the present work: 
\begin{itemize}
\item
The treatment of the meson--baryon interaction with three flavours is
completely analogous to the two-flavour case disussed before.
\item
The inclusion of fields with higher spin (photon, $\Delta$-resonance,
etc.) is straightforward. Their components are simply added to the bosonic
and fermionic sectors, respectively.
\item
The completion of the one-loop renormalization for the pion--nucleon
interaction up to $\cO(p^4)$ may be achieved by a suitable extension of 
(\ref{HBfluc}).
\item
For the analysis of fermionic bound states, the complete form (\ref{KP})
of the supermatrix operator must be used, as terms quartic in the fermion
fields are relevant in this case.
\item
In analogy to the mesonic case \cite{GLAP}, the super-heat-kernel
representation might also be useful for the finite part of the one-loop
functional with two external baryons.
\end{itemize}

\section*{Acknowledgements}
I am indebted to Gerhard Ecker, J\" urg Gasser, Joachim Kambor and Marc 
Knecht for helpful discussions and useful comments.

\medskip\medskip
\newcounter{zaehler}
\renewcommand{\thesection}{\Alph{zaehler}}
\renewcommand{\theequation}{\Alph{zaehler}.\arabic{equation}}
\setcounter{zaehler}{1}
\setcounter{equation}{0}

\newpage
\section*{Appendix}
\label{appdx}
I consider first the integrals
\beqa
I_n(v^2) := \int \dfrac{d^dl}{(2\pi)^d} \dfrac{e^{iv \cdot l}}{(l^2 +
i \epsilon)^n}
= f_n(d) (v^2)^{\frac{2n-d}{2}}
\label{In}
\eeqa
with an arbitrary four-vector $v_\mu$. The $f_n(d)$ are given by
\beq
f_1(d) = (-i)^{d-1} \dfrac{\Gamma(d-2)}{(4\pi)^{\frac{d-1}{2}}
\Gamma(\frac{d-1}{2})}
\eeq
and
\beq
f_n(d) = \dfrac{1}{2^{n-1} \, (n-1)! \, (d-2n) \ldots (d-4)} f_1(d)~,
\qquad
n =
2,3, \ldots
\eeq
The momentum space integrals occurring in (\ref{divergences}) are now 
obtained by differentiating (\ref{In}) a sufficient number of times with
respect to $v_\mu$ and
setting $v^2 = 1$ at the end:
\beqa
\int \dfrac{d^dl}{(2\pi)^d} \dfrac{e^{iv \cdot l}}{l^2}
&\stackrel{d \ra 4}{\longrightarrow}& \dfrac{i}{4 \pi^2}~, \\
\int \dfrac{d^dl}{(2\pi)^d} \dfrac{e^{iv \cdot l}}{l^2} l_\mu l_\nu
&\stackrel{d \ra 4}{\longrightarrow}& \dfrac{i}{2 \pi^2} \left(  g_{\mu
\nu} 
- 4 v_\mu v_\nu \right)~, \\
\int \dfrac{d^dl}{(2\pi)^d} \dfrac{e^{iv \cdot l}}{(l^2)^2} l_\mu
&\stackrel{d \ra 4}{\longrightarrow}& - \dfrac{v_\mu}{8 \pi^2}~, \\
\int \dfrac{d^dl}{(2\pi)^d} \dfrac{e^{iv \cdot l}}{(l^2)^2} l_\mu l_\nu
&\stackrel{d \ra 4}{\longrightarrow}& \dfrac{i}{8 \pi^2}
\left( g_{\mu \nu} - 2 v_\mu v_\nu \right)~, \\
\int \dfrac{d^dl}{(2\pi)^d} \dfrac{e^{iv \cdot l}}{(l^2)^2} l_\mu l_\nu
l_{\rho} 
&\stackrel{d \ra 4}{\longrightarrow}& \dfrac{1}{4 \pi^2}
\left[ -  \left( g_{\mu \nu} v_{\rho} + \ldots \right) + 4 v_\mu v_\nu
v_{\rho} \right]~, \\
\int \dfrac{d^dl}{(2\pi)^d} \dfrac{e^{iv \cdot l}}{(l^2)^3} l_\mu l_\nu
l_{\rho} 
&\stackrel{d \ra 4}{\longrightarrow}& \dfrac{1}{32 \pi^2}
\left[ -  \left( g_{\mu \nu} v_{\rho} + \ldots \right) + 2 v_\mu v_\nu
v_{\rho} \right]~, \\
\int \dfrac{d^dl}{(2\pi)^d} \dfrac{e^{iv \cdot l}}{(l^2)^3} l_\mu l_\nu
l_{\rho} l_{\sigma}
&\stackrel{d \ra 4}{\longrightarrow}& \dfrac{i}{32 \pi^2}
\left[ \left( g_{\mu \nu} g_{\rho \sigma} + \ldots \right) -  2 \left(
g_{\mu \nu} v_{\rho} v_\sigma + \ldots \right) \right. \nn
&& \qquad \quad \left. + 8 v_\mu v_\nu v_{\rho} v_\sigma \right]~, \\
\int \dfrac{d^dl}{(2\pi)^d} \dfrac{e^{iv \cdot l}}{(l^2)^4} l_\mu l_\nu
l_{\rho} l_{\sigma} l_\tau
&\stackrel{d \ra 4}{\longrightarrow}& \dfrac{1}{192 \pi^2}
\left[- \left( g_{\mu \nu} g_{\rho \sigma} v_\tau + \ldots \right) +  2
\left(
g_{\mu \nu} v_{\rho} v_\sigma v_\tau+ \ldots \right) \right. \nn
&& \qquad \quad \, \left. - 8 v_\mu v_\nu v_{\rho} v_\sigma v_\tau
\right]~. 
\eeqa
The dots indicate the necessary symmetrizations.

\end{document}